# IR LASER OPERATED PIEZOOPTICS IN PbTe:Ca CRYSTALS

M. Rezaee Roknabad[1], M. Mollaee[1*], M. Razavi[2], M. Mollai[2]

[1]Department of Physics (Thin Films Laboratory), Ferdowsi University of Mashhad, Iran.
[2]Department of Physics, Islamic Azad University, Mashhad Branch, Mashhad, Iran.

mmollaee@miners.utep.edu[*]

**Abstract**

IR induced piezo–optic effect (POE) in PbTe:Ca crystals was found under the influence of nanosecond - pulse $CO_2$ laser with a wavelength of 10.6μm. It was shown that addition of Ca leads to an increase of the POE tensor coefficient. This indicates the appearance of enhanced IR induced static dipole moments caused predominately by Ca impurities. Simultaneously, the variations in time kinetics for the POE in nanosecond time regime were explored. A substantial role of electron–phonon subsystem in the observed POE effect was demonstrated. The studies were done both for diagonal as well as off-diagonal POE tensor components.

## 1. Introduction

Lead telluride (PbTe) single crystals belong to semiconductors described by the general formula $A^{IV}B^{VI}$. Their physical and chemical properties have much in common: they have the same type of chemical bonds and they are isomorphous with cubic sodium chloride-type lattice [1]. Generally, PbTe is a nonstoichiometric compound ; therefore, it may be obtained with n- or p-type conductivity [2]. PbTe is particularly promising for high-ZT thermoelectric devices owing to its narrow band gap (0.31 eV at 300 K), face-centered cubic structure and large average exciton Bohr radius (~46 nm) provide strong quantum confinement within a large range of size. Theoretical calculations of the corresponding size effects are presented in [3, 4].

$Pb_{1-x}Ca_xTe$ has been grown for the first time using the technique of molecular beam epitaxy. Its index of refraction decreased with increasing Ca concentration for $x < 0.15$, as it is desired in a double heterojunction laser structure with Ca in the confining layers [5]. The electron and hole mobilities of $Pb_{1-x}Ca_xTe$ were comparable, and decreased roughly exponentially with calcium concentrations up to Ca content equal to about 0.15 [5]. The X-ray diffraction studies showed that the crystal lattice constant was almost independent of $x$. These results suggest that $Pb_{1-x}Ca_xTe$ may also be a useful material for long wavelength *($\lambda > 2\mu m$)* diode lasers [5].

The piezo-optic effect (POE) is generally closely related to acousto-optic efficiency of optical materials [6]. So it may be of a special interest for such kinds of studies. Moreover, mechanical stresses can cause an isotropic and inhomogeneous space distribution of the refractive indices. Their influence on the performance of optical waveguides have been observed in photoelectric devices [7] and the corresponding values of POE coefficient reported to range between $10^{-13}$ and $10^{-15}$ $m^2/N$ [8]. In this work for the first time it will be studied princiaplly new IR induced elastooticity.

## 2. Experimental

### 2.1. Specimen fabrication

Bridgman technique was used to grow PbTe:Ca single crystals which were synthesized in graphite crucibles in xenon ambient atmosphere (8MPa). The -grown crystals had a cylinder-like form with cylinder diameter and length equal to 10 and 60 mm, respectively. Their quality was controlled by the XRD method with Cu Kα source which unambiguously confirmed their single crystalline structure.

### 2.2. Non-linear optical measurements

The parametrically operated $CO_2$ laser received the laser generation wavelength at 10.6 μm with power densities varying within 0.5 – 1.5 $GW/cm^2$. The pulse duration was about 150 ns. The measurement set-up had two channels: photoinduced and pumping ones. These beams were temporarily shifted by delaying line. Additionally, a traditional Senarmont scheme was used for registration of the IR-induced birefringence [9].

Such short-time kinetics allowed us to avoid the sample overheating. The pumping laser beam has been scanned through the specimen's surface in order to eliminate inhomogeneous signal contributions. The measurements were carried out both for the transmitted as well as for the scattered light beams. Therefore, for every thickness (varied within the 200 – 450 μm) more than 40 measured points were statistically averaged  and scattered R lights intensities. The measurement devices both for the photoinduced and probing channels were temporarily synchronized. The absorption coefficients were evaluated taking into account sample's reflection and transparencies.

Light intensities for the time-dependent propagation were measured at different times $t$ of the pumping pulses with respect to the probing one. From the obtained system of equations towards $t$ we have evaluated the values of the time-delayed POE.

The average statistics over the sample surface was performed to avoid space non-uniformity in the sample distribution within the sample's surface.

### 3. Results and discussion

In Fig.1 dependences of the photoinduced POE diagonal coefficients versus pump–probe delaying time at different Ca content are presented.  It was established that the maximum of the POE tensor component achieves its maximal value of 2.5 $10^{-12}$ $m^2/N$ at pump–probe delaying time about 40-45 ns and power density 0.9 $GW/cm^2$. It is necessary to add that only diagonal $P_{22}$ tensor component is presented because the off-diagonal tensor components had the corresponding values at least one orders less. It is necessary to emphasize the non-monotonic features of the POE pump power density dependences.

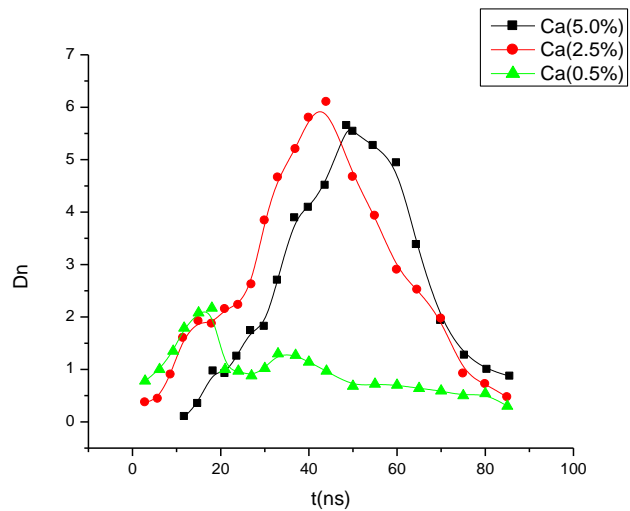

Fig.1. Photoinduced POE induced birefringence Dn diagonal tensor component versus pump-probe delaying time for the sample PbTe-Ca of different content at IR-induced power 0.9 $GW/cm^2$.

The second interesting fact consists in a temporary shift of the pump–probe POE maxima towards higher pump-probing times (from 20 to 50 ns) with increasing Ca content. One can see that there is a significant difference between specimens that possess Ca content equal to 0.5% and 2.5%. Additionally there is also obvious that there is no significant difference between 2.5% and 5% Ca content.  Both of them have the same maximum at the same pump-probing time. Moreover the observed time-delaying POE shift demonstrates asymmetric shape which is typical for decay of photoinduced carriers with   relatively low electron-phonon anharmonicities, so we can get a result in order to reach high value of Δn(changes of birefringence). So we need to use doping equal to about 2.5 %.

In Fig.2 are presented dependences of the photoinduced POE versus pump-probe delaying time at different IR-induced powers. One can see that maximum of the POE achieves maximal value of $3\ 10^{-12}$ m$^2$/N at pump-probe delaying times equal to  approximately 20 ps at 1 GW/cm$^2$. The maximum POE value is equal to $2.5\ 10^{-12}$ m$^2$/N and its value is decreased to $1.5\ 10^{-12}$ m$^2$/N. One can see that the POE coefficient increase linearly with increase of power density .Also it can be seen that with increase of photoinduced power density the slopes of curves increase and the peak becomes sharper. As we can see in the curve B maximum value becomes wide between 20 ps till 40 ps. All the curves reach the maximal value at almost the same time (about 20ps) and again they get together at the same point, about 60 ps.

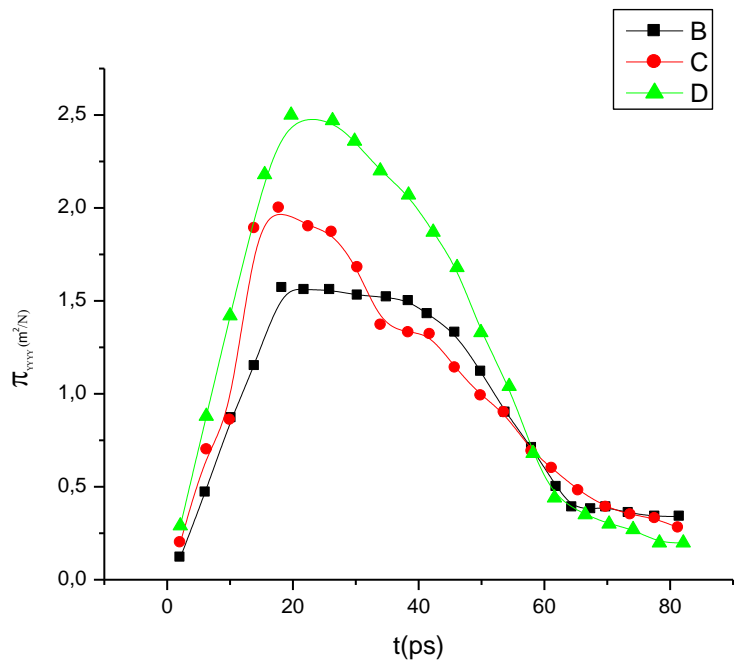

Fig. 2. Photo induced pump-probe POE kinetics for the sample at different photo inducing powers: B – power density 0.5   GW/cm$^2$ ; C - 0.8 GW/cm$^2$  D - 1GW/cm$^2$ . The POE coefficient should be Multiplied by $10^{-14}$.

Fig.3 presents a time decay of photo induced POE birefringence after switching off of the photo induced   power and at photo induced decay equal to about 80ps

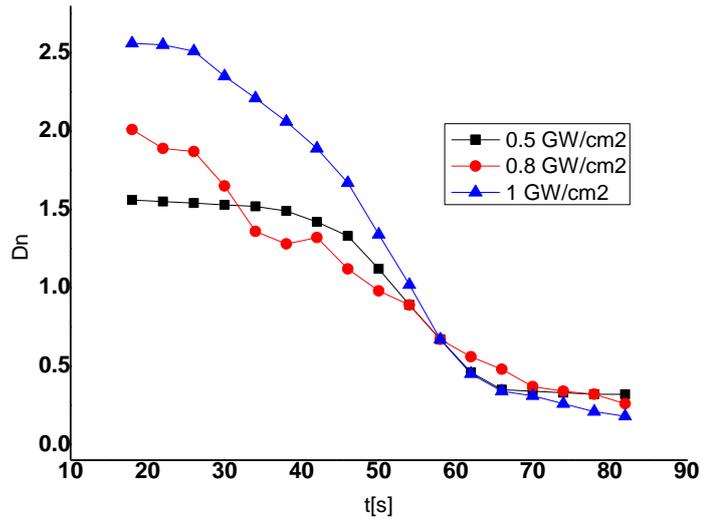

Fig. 3. IR-induced POE diagonal tensor decays for the samples photoinduced with different power.

The presented results demonstrate a principal role of the electron-phonon anharmonic interactions in the considered fourth order piezooptical tensors following the results obtained in the ref. 14-23. It is a consequence of IR-induced phonons effectively interacting with external laser beam. Moreover, they open a rare possibility to operate by the elastooptical and POE coefficient of materials using the IR laser beams.

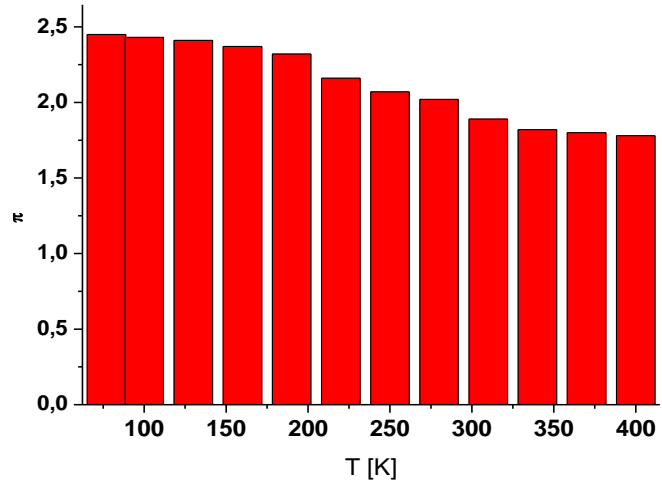

Fig. 4. Relative chagnes of diagonal photoinduced POE versus temperature at pump power density 1 GW/cm$^2$.

To study a contribution of the phonon subsystem we have studied temperature dependences of the photo induced POE for optimal photo inducing power i.e. at 1 GW/cm$^2$. One can see that within the temperature region 220K 280K there is observed significant changes of the piezooptical coefficients reflecting changes of the corresponding elastooptical constants, which indicate on substantial contribution of the phonons to the observed effect.

## 4. Conclusions

Experimental studies of IR induced elasto (piezo-)-optics effect in PbTe:Ca crystals were performed under influence of nanosecond $CO_2$ laser wavelength 10.6μm. It was established that addition of Ca leads to an increase of the POE coefficient up to $2.5 \cdot 10^{-12} m^2/N$. This one indicates on appearance of enhanced IR induced static dipole moments caused by defects. Simultaneously changes of time kinetics of the POE in nanosecond time regime were studied. A substantial role of electron–phonon subsystem in the observed EOE is shown. The studies were done both for diagonal as well as off-diagonal tensor components. The presented results demonstrate a principal role of the electron-phonon anharmonic interactions in the considered fourth order piezooptical tensors. One can see that the maximum of the POE tensor coefficients achieves its maximal value of $3 \cdot 10^{-12} m^2/N$ at pump-probe delaying time equal to approximately 20ps at $1 GW/cm^2$.